\documentclass{epl}

\title{Logarithmic Growth Dynamics in Software Networks}
\author{Sergi Valverde\inst{1} \and Ricard V. Sol\'e\inst{1,2}}
\institute{
  \inst{1} ICREA-Complex Systems Lab - Universitat Pompeu Fabra, Dr. Aiguader 80, 08003 Barcelona, Spain \\
  \inst{2} Santa Fe Institute - 1399 Hyde Park Road, New Mexico 87501, USA
}


\begin{document}

\maketitle

\begin{abstract}
In a recent paper, Krapivsky and Redner\cite{Lognets} proposed a new growing
network model with new nodes being attached to a randomly selected node, as
well to all ancestors of the target node. The model leads to a sparse graph
with an average degree growing logarithmically with the system size. Here we present
compeling evidence for software networks being the result of a similar class of
growing dynamics. The predicted patternd of network growth, as well as the
stationary in- and out-degree distributions are consistent with the model. Our 
results confirm the view of large-scale software topology being generated through 
duplication-rewiring mechanisms. Implications of these findings are outlined.
\end{abstract}

\section{Introduction}

\begin{figure}
\onefigure[width=0.95\textwidth]{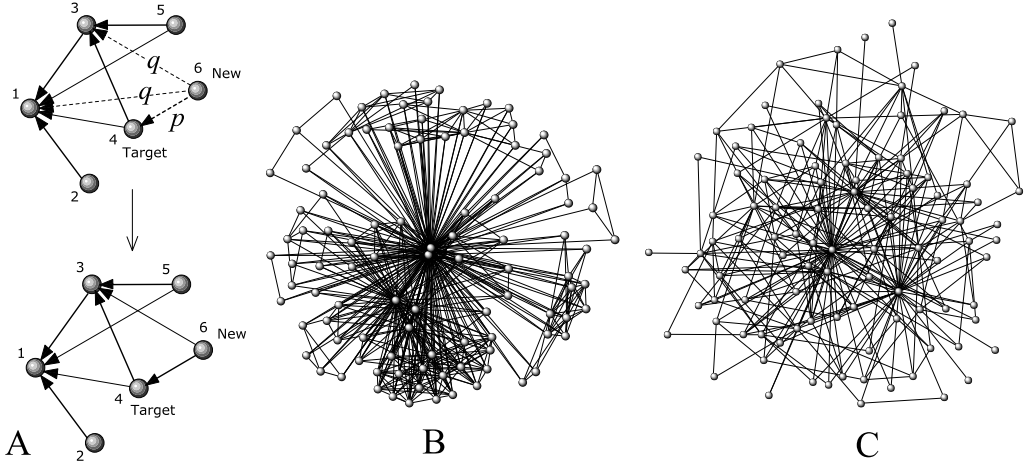}
\caption{ (A) Illustration of the copying rule used in the network growth model.
Each node is labeled with a number indicating its age (number one 
is the oldest). In the figure, new node $v_6$ attaches to target node $v_4$
with probability $p$. This new node inherits every link from target node (dashed links), 
with probability $q$. (B) Synthetic network obtained with the GNC model with 
$N=100$, $m=1$, $p=1$ and $q=1$. (C) Synthetic network obtained with the 
GNC model with $N=100$, $m=4$, $p=0.25$ and $q=0.25$. These networks have 
a scale-free in-degree distribution and an exponential out-degree distribution. }
\label{f.1}
\end{figure}

The structure of many natural and artificial systems can be depicted 
with networks. Empirical studies on these networks have revealed that
many of them display a heterogenous degree distribution 
$p(k) \approx k^{-\gamma}$, where few nodes (hubs) have a large number of 
connections while the majority of nodes have one or two links \cite{Dorogovtsev}. 
The existence of hubs has been related to multiplicative effects affecting 
network evolution \cite{Barabasi}. Such topological patterns have been
explained by a number of mechanisms, including preferential attachment rules 
\cite{Caldarelli} and network models based on simple rules of node 
duplication \cite{duplication}. A very simple approach is given by the
growing network model with copying (GNC)\cite{Lognets}. The network grows by introducing a single node 
at a time. This new node links to $m$ randomly selected target node(s) with probability $p$ 
as well to all ancestor nodes of each target, with probability $q$ (see fig.~\ref{f.1}). 
The discrete dynamics follows a rate equation \cite{Lognets}

\begin{equation}
L(N + 1) = L(N) + \frac{m}{N}\left\langle {\sum\limits_\mu  {\left( {p + qj_\mu  } \right)} } \right\rangle 
\label{disc}
\end{equation}

where $L$ and $N$ are the number of links and nodes, respectively. The second term in
the right-hand side describes the copying process, where the average number of links
added is given by $p + qj_\mu$ . The $\mu$ index refers to the node $\mu$, to 
be selected uniformly from among the $N$ elements. Assuming a continuum approximation, 
the number of links is driven by the following differential equation:

\begin{equation}
\frac{{dL}}{{dN}} = mp + mq\frac{L}{N}
\label{e.1}
\end{equation}

The asymptotic growth of the average total number of links depends on the extent 
of copying defined by the product $mq$. In particular, logarithmic growth is recovered
when $mq=1$ and $L(N) = mpN\log N$. This corresponds to a marginal 
situation separating a domain of linear growth ($mq<1$) to a domain of 
exponential growth ($2>mq>1$). Interestingly, for $mq=1$ the GNC model predicts a 
power-law in-degree distribution $P_i(k) \approx k^{-\gamma_i}$ with exponent 
$\gamma_i = 2$ and an exponential out-degree distribution $P_o(k)$, independently 
of copying parameters. Actually, their derivation for the in-degree distribution
can be generalized for arbitrary $q$ and $p$ values, leading to a scaling
law $P_i(k) \approx k^{-2}$ for the parameter domain of interest. In ref. \cite{Lognets} 
the authors showed that the GNC model seems to consistently explain the 
patterns displayed by citation networks. Here, we show that a GNC model is also 
consistent with the evolution of software designs, which also display the 
predicted logarithmic growth. 

\section{Software Networks}

One of the most important technological networks, together with the Internet
and the power grid, is represented by a broad class of software maps. Software
actually pervades technological complexity, since the control and stability 
of transport systems, Internet and the power grid largely rely on sotfware-based 
systems. In spite of the multiplicity and diversity of objectives and functionalities addressed
by software projects, we have pointed out the existence of strong universals
in their internal organization \cite{Europhys}. Computer programs are often decomposed in a collection of
text files written in a given programming language. In this paper, we will study computer
programs written in C or C++\cite{cpp}. Program structure can be recovered from the 
analysis of program files by means of a directed network representation. 
In a software network, software entities (files, classes, functions or instructions) 
map onto nodes and links representing syntactical dependencies \cite{Europhys}. 
Class graphs (also called 'logical dependency graph' \cite{Lakos}) are a 
particular class of software networks that has been shown to be small-world 
and scale-free network with an exponent $\gamma \approx 2.5$  \cite{Europhys, IEEE, Myers}. 
 Interestingly, the frequencies of common motifs displayed 
in class graphs can be predicted by a very simple duplication-based model of 
network growth \cite {PREMotifs}. This result indicates
that the topology of technological designs, in spite of being planned and problem-dependent,
might actually emerge from common, distributed rules of tinkering \cite{Complexity}.
In the following,  we provide further evidence for the importance of duplication 
processes in the evolution of software networks. 

Here, we study a new class of software networks. We 
use the so-called 'include graph' (or 'physical dependency graph' in \cite{Lakos}) 
$G = (V,E)$ where $v_i \in V$ is a program file and a directed 
link $(v_i, v_j) \in E$ indicates a (compile-time) dependency
between file $v_i$ and $v_j$. 
In C and C++, such dependencies are encoded with the 
keyword "\#include" followed by the name of the refereed source 
file \cite{Lakos}.  In order to
recover the include graph, we have implemented a network
reconstruction algorithm that analyses the contents of all files 
in the software project looking for this reserved keyword. Every
time this keyword is found in a file $v_i$, the name of the 
refereed file $v_j$ is decoded and a new link $(v_i, v_j)$ is added.
No other information is considered by the network 
reconstruction algorithm. Notice that the include network is unweighted
because it makes no difference to include the same file twice.

In this paper, we investigate the structure and evolution of software maps by looking at 
their topological structure and the time series of 
aggregate topological measures, such as number 
of nodes $N(t)$, number of links $L(t)$ or average degree $k(t) = L(t)/N(t)$. 
It is worth mentioning that the number of nodes in a include graph coincides with 
the number of files in the software project, which is often used 
as a measure of project size. 

\begin{figure}
\onefigure[width=.98\textwidth]{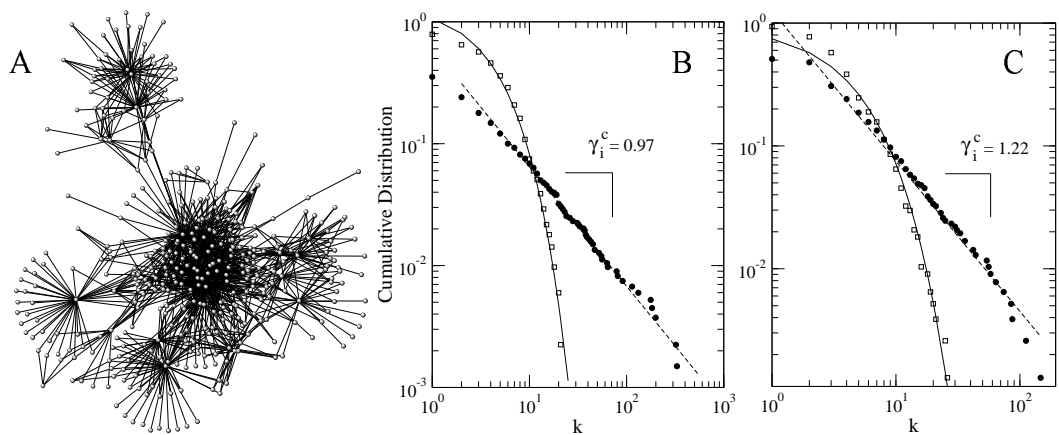}
\caption{ 
(A) Largest connected component of the XFree86 include network 
at 15/05/1994 (with $N=393$) displays scale-free behavior (see text). In (B),
the cumulative distributions $P_{i>}(k)$ and $P_{o>}(k)$ are shown 
for a more recent version of the XFree86 include network with $N=1299$
(not shown here). 
The power-law fit of the in-degree distribution yields  
$P_i(k) \sim k^{-\gamma_i^c-1}$ with $\gamma_i^c = 0.97 \pm 0.01$ 
while the out-degree distribution is exponential. In (C) we can notice 
similar features for the in-degree and out-degree distributions of the
Aztec include network at 29/3/2003. For this system, the power-law
fit of the in-degree distribution yields an exponent $\gamma_i^c =
1.22 \pm 0.03$. }
\label{f.2}
\end{figure}


Software maps typically display asymmetries in their in-and out-degree 
distributions \cite{Myers,IEEE} although the origins of such asymmetry remained unclear. 
Notice how the out-degree and in-degree distributions of real include networks 
are quite similar to the corresponding distributions
 obtained with the GNC model (see previous section). The in-degree and out-degree 
distributions for the largest component of two different systems are shown 
in fig.~\ref{f.2}B and fig.~\ref{f.2}C,  where we have used the
cumulative distributions $P_ >  (k) = \int\limits_k^\infty  {P(k)dk}$.
In both cases, in-degree distributions display scaling $P_{i}(k) \approx k^{-\gamma_{i}}$, where the
estimated exponent is consistent with the prediction from the GNC
model, whereas out-degree distributions are single-scaled (here the average value for 
the systems analysed is $\left\langle {\gamma _i } \right\rangle = 2.08 \pm 0.04$ \cite{website}). As shown in the 
next section, these stationary distributions result from a logarithmic growth 
dynamics consistent with the GNC model.


\section{Software Evolution}


Although an extensive literature on software evolution exists (see for 
example \cite{Belady76,Godfrey}, little 
quantitative predictions have been presented so far. Most studies are actually 
descriptive and untested conjectures about the nature of the constraints 
acting on the growth of software structures abound. It is not clear, 
for example, if the large-scale patterns are due to external constraints, 
path-dependent processes or specific functionalities. In order to 
answer these questions, we have compared real software evolution 
with models of network growth, where software size is measured as the number 
of nodes in the corresponding include graph. In this context, the assumptions 
of the GNC model are consistent with observations claiming that 
code cloning is a common practice in software development \cite{Godfrey}. Indeed, 
comparison between real include graphs and those generated with the GNC model 
suggests the extent of copying performed during software evolution is a 
key parameter that explains the overall pattern of software growth. Such a 
situation has been also found in class diagrams \cite{PREMotifs}.

\begin{figure}
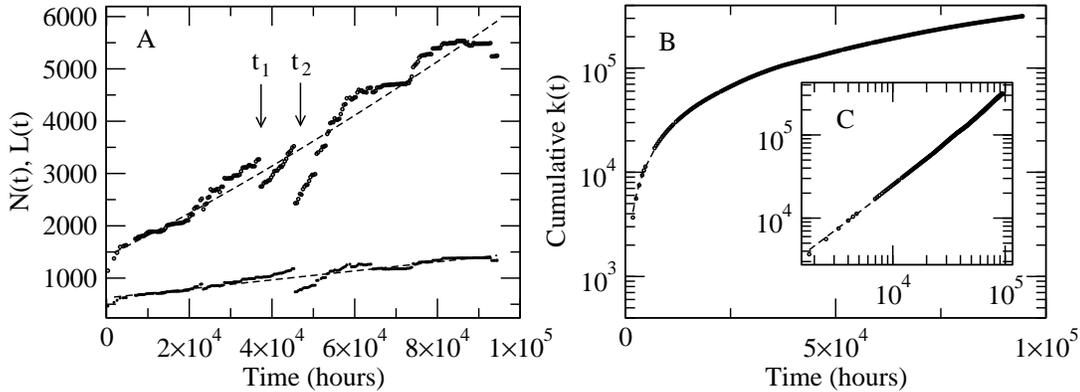

\onefigure[width=1.0\textwidth]{Xfree86_evo4.eps}
\caption{(A) The top curve shows the comparison between the time evolution of number 
of links $L(t)$ in XFree86 between
$16/05/1994$ and $01/06/2005$ (points) and the prediction of eq.~(\ref{e.4}) (dashed line). 
In the bottom curve we compare the time evolution of system size $N(t)$ and 
its linear fitting $N(t)=N_0+at$ (dashed line). We observe an anomalous growth pattern followed 
by a discontinuity (here indicated as $t_1$ and $t_2$) in $L(t)$. Notice how $t_2$ signals
a discontinuity both in $L(t)$ and $N(t)$, while discontinuity $t_1$ only takes place in $L(t)$. 
(B) Comparison between time evolution of the cumulative average degree in XFree86 during the
same time period as in (A) and the analytic prediction of eq.~(\ref{e.5}). (C) The inset shows 
the same data as in (B) but in a double logarithmic plot. The fitting
 parameters are: $N_0 = 622.17 \pm 10.92$, 
$a = 0.0086 \pm 0.0002$, $L_0 = 1419.8 \pm 4.1$, 
and $mp = 2.20 \pm 0.01$. Time is measured in hours. }
\label{f.3}
\end{figure}

The growth dynamics found in include graphs is logarithmic (see fig.~\ref{f.3}A) 
thus indicating that we are close to the $mq=1$ regime. Indeed, the sparseness 
seen in software maps is likely to result from a compromise between having enough 
dependencies to provide diversity and complexity (which require more links) 
and evolvability and flexibility (requiring less connections). Here we 
have uneven, but detailed information of the process of software building.
In this context, different software projects developments display specific 
patterns of growth. Specifically, the number of nodes $N$ grow with time
following a case-dependent functional form $N = \phi(t)$. Using 
$dL/dt=(dL/dN)(d\phi/dt)$, we have from, eq.~(\ref{e.1}),
\begin{equation}
\frac{{dL}}{{dt}} = \left[ {mp + mq\frac{L}{{\Phi (t)}}} \right]\dot \Phi ^{ - 1} 
\end{equation}
with a general solution
\begin{equation}
L(t) = e^{mq\int ( \Phi \dot \Phi )^{ - 1} dt} \left[ {mp\int {e^{ - mq\int ( \Phi \dot \Phi )^{ - 1} dt} } \dot \Phi ^{ - 1} dt + \Gamma } \right]
\label{solution}
\end{equation}
where $\Gamma$ is a constant. Using a linear law growth (which is not uncommon in 
software development), i.e. $N(t)= N_0 + at$, and assuming $mq=1$, we have,

\begin{equation}
L(t) = \left( {N_0  + at} \right)\left[ {mp\log \left( {\frac{{N_0  + at}}{{N_0 }}} \right) + \frac{{L_0 }}{{N_0 }}} \right]
\label{e.4}
\end{equation}

However, typical time series of $L(t)$ in real software evolution is subject to fluctuations (see
fig.~\ref{f.3}A).  In order to reduce the impact of fluctuations we use the cumulative average degree
$K(t) = \int \limits_0^t {(L/N)dt}$, instead. Assuming the number of nodes growths linearly in time, we obtain:

\begin{equation}
K(t) = \frac{{mp(N_0  + at)}}{a}\left[ {\log \left(\frac{{N_0  + at}}{{N_0 }} \right) - 1} \right] + \frac{{L_0 }}{{N_0 }}t + \frac{{mpN_0 }}{a}
\label{e.5}
\end{equation}

The above expressions can be employed to estimate the parameters $L_0$ and $mp$ describing
the shape of the logarithmic growth of number of links L(t) and the parameters $N_0$ and $a$
controlling the linear growth of the number of nodes N(t).  We used the following fitting procedure. 
For each software project, we have recovered a temporal sequence 
$\left\{ {G_t  = (V_t ,E_t )} \right.\left. {\left| {0 \le t \le T} \right.} \right\}$
of include networks corresponding to different versions of the software project. Time
is measured in elapsed hours since the first observed project version (which can or cannot
coincide with the beginning of the project). This temporal sequence describes the evolution 
of the software project under study. From this sequence, we compute the evolution of 
the number of nodes $n_0, n_1, n_2, ... ,n_T$, the evolution of the number of 
links $l_0, l_1, l_2, ... , l_T$ and the evolution of the average degree 
$k_0, k_2, ... , k_i=l_i/n_i, ... , k_T$.  In general, available data is a 
partial set of records of development histories and often misses the initial 
project versions corresponding to the early evolution. Then, $t_0  \ne 0$ and this explains why the 
initial observations for $n_0$ and $l_0$ are higher than expected. However, we have rescaled 
time so the first datapoint corresponds to zero. We have collected partial \footnote{Actually, 
these datasets constitute a coarse sampling of the underlying process of software change. 
Collecting software evolution data at the finest level of resolution requires a monitoring 
system that tracks automatically all changes made by programmers. Instead, it is often the
programmer who decides when a software register is created. The issue of fine-grained
sampling is an open research question in empirical software engineering that deserves
more attention. These limitations preclude us a more direct testing of the GNC model.}
 evolution registers for seven different projects (relevant time period is in parenthesis): 
XFree86 (16/5/94 - 1/6/05), Postgresql (1/1/95 - 1/12/04), DCPlusPlus (1/12/01 - 15/12/04),
TortoiseCVS (15/1/01 - 1/6/05), Aztec (22/3/01 - 14/4/03), Emule (6/7/02 - 26/7/05) and 
VirtualDub (15/8/00 - 10/7/05) \cite{website}. The full database comprises 557 include 
networks (see table ~\ref{t.2}). 

Then, we proceed as follows. First, for each software project, its time series for the number 
of nodes is fitted under the assumption of linear growth, i.e. $N(t)= N_0 + at$, 
and thus yielding $N_0$ (initial number of nodes) and $a$ (rate of addition of new files). In table ~\ref{t.2} , we 
can appreciate that the majority of projects growth at a rate $a$ proportional 
to $10^{-3}$ files/hour while two medium size projects (Aztec and Emule) 
actually grow by an order of magnitude faster. Next, we compute the
time series of cumulative average degree $K(t)$ by integrating numerically the
sequence of $k_t$ values. This new sequence will be fitted with eq.~\ref{e.5}
 in order to estimate the parameters $L_0$ (initial number of links)
and the product $mp$ controlling the extent of duplication.

\begin{table}
\caption{Predictions of eq.~(\ref{e.5}) for different systems.}
\label{t.2}
\begin{center}
\begin{tabular}{lccccc}
Project   &     $a$                    & $N_0$          & $mp$ &          $L_0$ &           $T$ \\
XFree86 	& $0.0086 \pm 0.0001$ 	 & $622.17 \pm 10.92$ & $2.20 \pm 0.01$ & $1419.80 \pm 4.09$ &  243\\
Postgresql 	& $0.0066 \pm 0.0002$    & $601.42 \pm 11.35$ & $1.78 \pm 0.05$ & $243.89 \pm 8.46$  & 31\\
DCPlusPlus 	& $0.004 \pm 0.0001$     & $101.51 \pm 2.42$  & $0.70 \pm 0.03$ & $338.96 \pm 1.30$  & 74 \\
TortoiseCVS & $0.0057 \pm 0.0001$ 	 & $97.57 \pm 2.62$   & $1.59 \pm 0.02$ & $105.76 \pm 1.58$  & 107\\
Aztec		& $0.026 \pm 0.002$ 	 & $205.12 \pm 22.17$ & $0.97 \pm 0.03$ & $622.61 \pm 4.77$  & 14 \\
Emule		& $0.016 \pm 0.0006$	 & $98.01 \pm 6.37$   & $1.65 \pm 0.11$ & $223.80 \pm 9.34$  & 54 \\
VirtualDub  & $0.0079 \pm 0.0004$    & $167.04 \pm 12.44$ & $1.34 \pm 0.05$ & $381.50 \pm 5.16$  & 35 
\end{tabular}
\end{center}
\end{table}

In fig.~\ref{f.3}B we show the result of the previous fitting procedure to the 
time series of cumulative average degree $K(t)$ in XFree86, a popular and freely 
re-distributable open source implementation of the X Windows System 
\cite{website}. As shown in the figure, the agreement 
between theory and data is very good. We have validated the same 
logarithmic growth pattern in the evolution of other software 
systems (see table ~\ref{t.2}). In particular, we provide a prediction for the 
average number of links to target nodes, $mp$, which is found to be small. This is 
again expected from the sparse graphs that are generated through the growth process. 

Together with the overall trends, we also see deviations from the logarithmic 
growth followed by reset events. 
In fig.~\ref{f.3}A we can appreciate a pattern of discontinuous software growth
in the number of links $L(t)$ for XFree86. The time interval delimited by 
$t_1$ and $t_2$ is the signature of a well-known major 
redesign process that enabled 3D rendering capabilities in XFree86.
This new feature of XFree86 was called Direct Rendering Infrastructure (DRI).
Development of DRI is cleary visible in the time series of $L(t)$. 
At $t_1$ (i.e., August of 1998) the design of DRI 
was officially initiated and the event $t_2$ (i.e., July of 1999) corresponds 
to the first public release of the DRI technology (i.e., DRI 1.0)
\cite{Xfree86}. A careful look at the time series $L(t)$ shows that before the 
discontinuities (indicated by $t_1$ and $t_2$), some type of precursor
patterns were detectable.

The above example suggests how deviations from the logarithmic growth pattern 
can predict future episodes of costly internal reorganization 
(so called refactorings \cite{Refactor}). In XFree86, the integration of DRI was a
costly redesign process characterized by an exponential growth pattern 
in the number of links L(t). This accelerated growth pattern starts at $t_1$ and
finishes in a clearly visible discontinuity (indicated here by $t_2$)
 that signals a heavy removal of links. After $t_2$ we observe a 
pattern of fast recovery eventually returning to the 
logarithmic trend described by eq.~(\ref{e.4})(dashed lines in fig fig.~\ref{f.3}A). 
Such type of reset pattern has been also found in economic fluctuations in the 
stock market \cite{Sornette}. This trend needs to be explained and might actually 
result from conflicting constraints leading to some class of marginal equilibrium 
state. This is actually in agreement with the patterns of activity change 
displayed by the community of software developers (unpublished results) 
which also exhibits scale-free fluctuations.

\acknowledgments
We thank our colleague J. F. Sebastian for useful suggestions.
This work has been supported by grants BFM2001-2154 and by the EU within the 6th 
Framework Program under contract 001907 (DELIS) and by the Santa Fe Institute.

\end{document}